\documentstyle[aps,prd]{revtex}
\begin{document}

\newcommand{\MF}{{\large{\manual META}\-{\manual FONT}}}
\newcommand{\manual}{rm}
\newcommand\bs{\char '134 }
\title{Renormalization of Black Hole Entropy}
\author{{\rm Sang Pyo} Kim$^a$ \thanks{Electronic mail: sangkim@knusun1.kunsan.ac.kr},
{\rm Sung Ku} Kim$^b$ \thanks{Electronic mail: skkim@theory.ewha.ac.kr},
{\rm Kwang-Sup} Soh$^c$ \thanks{Electronic mail: kssoh@phyb.snu.ac.kr},
{\rm and} {\rm Jae Hyung} Yee$^d$ \thanks{Electronic mail: jhyee@phya.yonsei.ac.kr}}
\address{$^a$ Department of Physics, Kunsan National University,
Kunsan 573-701 \\
$^b$ Department of Physics, Ewha Womans University,
Seoul 120-750\\
$^c$ Department of Physics Education, Seoul National University,
Seoul 151-742\\
$^d$ Department of Physics and Institute for Mathematical
Sciences, Yonsei University, Seoul 120-749}
\maketitle

\begin{abstract}
We review the renormalization of one-loop  effective action for gravity
coupled to a scalar field and that of the Bekenstein-Hawking entropy
of a black hole plus the statistical entropy of the
scalar field. It is found that the total entropy of the black hole's
geometric entropy and the statistical entropy
yields the renormalized Bekenstein-Hawking area-law
of black hole entropy only for even
dimensional Reissner-N\"{o}rdstrom (Schwarzschild) black  holes.
We discuss the  problem of
the microscopic origin of black hole entropy in connection
with the renormalization of black hole entropy.
\end{abstract}

\section{Introduction}

One of the mysterious and challenging phenomena
in general relativity is black hole physics.
Black holes formed from the gravitational  collapse of massive  stars
are  characterized by mass $M$, angular momentum $J$ and charge $Q$,
regardless of the process of their formation.
Classically a black hole prevents any particle from crossing its event horizon
from the interior, and  all the particles falling into it
eventually run into the singularity. Thus black holes are classically black.
On the other hand, quantum mechanically the black hole emits a
thermal spectrum characterized by the Hawking temperature
and has the Bekenstein-Hawking entropy \cite{bekenstein,hawking}.

The Bekenstein-Hawking entropy of black hole, also known as the area-law,
is given by one quarter of the area of the event horizon of the black hole.
The black hole entropy is
a quantity entirely determined by the geometry of black hole. However,
in statistical mechanics the entropy of a system is determined by
the number of microscopically indistinguishable states
available to the system.
Thus the Bekenstein-Hawking entropy is lacking in the explanation of the
microscopic origin of entropy. As an attempt to understand the microscopic origin
of black hole entropy t' Hooft introduced a brick wall model, in which
a scalar field is confined to a spherical shell enclosing the event horizon
and when one chooses a thickness having the geometric meaning
of an invariant distance from the event horizon,
the statistical entropy of the scalar field computed in terms of the
thickness of the brick wall gives the correct black hole entropy \cite{thooft}.
This idea was more elaborated by considering the renormalization of the
total entropy of the black hole's geometric entropy and the statistical entropy of the
scalar field in the black hole background \cite{susskind}. The renormalization
scheme of black hole entropy was realized for Reissner-N\"{o}rdstrom (RN)
black hole minimally coupled to a massive scalar field in four
\cite{demers} and six dimensions \cite{kim1}.

In this talk we revisit the problem of the renormalization of
black hole entropy for black holes coupled to
a massive scalar field. By extending the previous result \cite{kim1,kim2},
we show explicitly that the renormalization of
the black hole's geometric entropy plus the statistical entropy of the
scalar field leads correctly to the renormalized Bekenstein-Hawking black
hole entropy only for all the even dimensional RN black holes.

The organization of this talk is as follows. In Sec. II we review
the Bekenstein-Hawking black hole entropy for a four-dimensional RN
black hole and the problem of the microscopic origin of black hole entropy.
In Sec. III we derive the Hawking temperature for the $D$-dimensional
RN black holes. In Sec. IV we apply the 't Hooft brick wall model to
$D$-dimensional RN black holes to explain the black hole entropy
in terms of the statistical entropy. In Sec. V we review the renormalization
of the one-loop effective action for gravity minimally coupled to
a massive scalar field for the even dimensional RN black holes.
Finally in Sec. VI we show explicitly the renormalization of the black
hole entropy plus the statistical entropy for all the even dimensional
RN black holes. We discuss the possible role of supersymmetry in the
black hole entropy problem.
\section{Bekenstein-Hawking Black Hole Entropy and Statistical
Entropy}
It was discovered by  Bekenstein and Hawking  that a black  hole has
a complete analog
with a thermal system with the Hawking temperature and the
Bekenstein-Hawking entropy.
A four-dimensional RN black hole has the temperature and entropy
\begin{eqnarray}
T_4 = \frac{1}{ \pi r_+} \Bigl(1 - \frac{r_-}{r_+} \Bigr),
~~~~S_4 = \frac{A_4}{4 G},
\end{eqnarray}
where $A_4 = 4  \pi r_+^2$ is the  area of the event horizon of the
RN black hole with the metric
\begin{eqnarray}
ds^2 &=& - \Bigl( 1 - \frac{r_-}{r} \Bigr) \Bigl(1 - \frac{r_+}{r} \Bigr)
dt^2 \nonumber \\
&&~~+ \frac{dr^2}{\Bigl(1 - \frac{r_-}{r} \Bigr) \Bigl( 1 - \frac{r_+}{r}
\Bigr)} + r^2 d \Omega_2^2,
\end{eqnarray}
where $r_+$ and $r_-$ are the outer and inner event horizons, respectively.
Use of geometrodynamical units $c = \hbar =  k_B = 1$ is made, but  $G$ will be kept  whenever
necessary.

We follow the beautiful derivation  by Gibbons and Hawking \cite{gibbons}  of the geometric
entropy of the four-dimensional RN black hole using the Euclidean path integral.  The partition
function for gravity plus matter typically represented by $\phi$
is given by
\begin{equation}
{\cal Z}_4 = e^{- \beta F_4} = \int {\cal D} [g] {\cal D} [\phi] e^{- {\cal I}_E [ g, \phi]}
\end{equation}
where ${\cal I}_E$ is the Wick rotation of the Einstein-Hilbert action
\begin{equation}
\hspace*{-5pt}
{\cal I}_{EH}  = \frac{1}{16  \pi G}  \int_{\cal M}   d^4x \sqrt{- g}  R +  \frac{1}{8 \pi   G}
\int_{\partial {\cal M}} d \Sigma [K],
\end{equation}
where $K = g^{ab}K_{ab}$, $K_{ab}$  being the second
fundamental form  on $\partial {\cal
M}$, and $[K] = K_{\vert \partial {\cal M}} - K_{\vert {\rm flat space}}$.
For the RN black hole, the bulk contribution vanishes
due to the on-shell condition (being a solution
of the Einstein equation), and only the surface contribution survives
to yield
\begin{equation}
{\cal I}_{EH} =  \frac{1}{8 \pi G} \frac{\partial}{\partial n}
\int_{\partial {\cal M}} d \Sigma.
\end{equation}
For the RN black hole the Einstein-Hilbert action is evaluated
\begin{equation}
{\cal I}_{EH} = i \frac{\pi r_+^2}{G}.
\end{equation}
>From the partition function one finds the Helmholtz free energy
\begin{equation}
F_4 = \frac{{\cal I}_E}{\beta} = - \frac{\pi r_+^2}{G \beta},
\end{equation}
and the entropy
\begin{equation}
S_4 = \beta^2 \frac{\partial F_4}{\partial \beta} = \frac{A_4}{4 G}.
\label{bk entropy}
\end{equation}
Eq. (\ref{bk entropy}) is the famous Bekenstein-Hawking black hole
entropy. Note that the Bekenstein-Hawking entropy is determined entirely
by the geometry of black hole itself.

On the other hand, in statistical mechanics
the entropy of a system is defined by the  number of microscopically
indistinguishable states $N$:
\begin{equation}
S = \ln (N).
\end{equation}
For a system in a thermal equilibrium the entropy is also given by
\begin{equation}
S = - {\rm Tr} \Bigl[ \hat{\rho} \ln (\hat{\rho}) \Bigr]
\end{equation}
where $\hat{\rho}$ is the density operator
\begin{equation}
\hat{\rho} = \frac{1}{Z} e^{- \beta \hat{H}}.
\end{equation}
By evaluating the partition function
\begin{eqnarray}
Z &=& {\rm Tr} \Bigl[e^{- \beta \hat{H}} \Bigr] \nonumber\\
  &\equiv& e^{ - \beta F},
\end{eqnarray}
one finds the entropy as before
\begin{equation}
S = \beta^2 \frac{\partial F}{\partial \beta}.
\end{equation}

Thus one may raise a question about the microscopic origin of
black hole entropy.
The area-law of Bekenstein-Hawking  black hole entropy
is rooted entirely on  the
geometry of black  hole,  whereas  the  entropy in  statistical
mechanics  counts  the  number  of microscopically
indistinguishable states available  to the system.
Then, where does the  entropy of black  hole reside?
There  have recently  been
several different approaches toward this problem of the
microscopic origin of black hole entropy.
Among these  string theory   seems  to  explain
the   correct number   of  string  states
$e^{\frac{A}{4G}}$ \cite{horowitz}.
However, one of the problems with which string theory confronts now is
to explain the topology of black holes from the spacetime of strings.

\section{Entropy of D-Dimensional RN Black Holes}

The $D$-dimensional RN black hole has the metric \cite{myers}
\begin{equation}
ds^2 = - \Delta (r) dt^2 + \Delta^{-1} (r) dr^2 + r^2 d \Omega_{D - 2}^2,
\label{n bh}
\end{equation}
where
\begin{equation}
\Delta (r) = \Biggl(1 - \Bigl( \frac{r_-}{r} \Bigr)^{D -3} \Biggr)
\Biggl(1 - \Bigl( \frac{r_+}{r} \Bigr)^{D -3} \Biggr),
\end{equation}
where $r_+$ and $r_-$ are the outer and inner event horizons given by
\begin{equation}
\hspace*{-15pt}
r_{\pm} =  \Biggl[ \frac{4 \Gamma \Bigl((D-1) \frac{1}{2} \Bigr)}{ (D -2)
\pi^{(D -3)/2}} \Bigl( M \pm \sqrt{M^2 - Q^2} \Bigr) \Biggr]^{\frac{1}{D -3}}.
\end{equation}

For the nonextremal black holes we can find the coordinate transformation
that removes the apparent singularity of the event horizon.
With a transformation of the form
\begin{equation}
r = r_+ + c \rho^q,
\end{equation}
the two dimensional metric near the event horizon becomes
\begin{equation}
- \Delta (r) dt^2  + \frac{dr^2}{\Delta (r)}
\simeq
- \frac{(D-3) c (1-u)}{r_+} dt^2
+ \frac{c q^2 r_+}{(D-3) (1 -u)} \rho^{q-2} d\rho^2 ,
\end{equation}
where $u = \Bigl(\frac{r_-}{r_+} \Bigr)^{D-3}$.
By choosing the undetermined parameters
\begin{equation}
q = 2,~~~~
c = \frac{(D-3) (1-u)}{ 4 r_+},
\end{equation}
and by introducing a new coordinate
\begin{equation}
\theta = \frac{(D-3) (1-u)}{2 r_+} t,
\end{equation}
we transform the black hole metric up to the order of ${\cal O} (\rho^2)$
\begin{eqnarray}
ds^2 &=& - \rho^2 \Biggl[1 + \frac{(D-3)}{4 r+^2}
\Bigl( (D-3) u - \frac{(D-2)(1-u)}{2} \Bigr) \rho^2 \Biggr] d \theta^2
\nonumber\\
&&~ + \Biggl[1 - \frac{(D-3)}{4 r+^2}
\Bigl( (D-3) u - \frac{(D-2)(1-u)}{2} \Bigr) \rho^2 \Biggr] d \rho^2
+ r_+^2 \Biggl[1 + \frac{(D-3) (1-u)}{4 r_+^2} \rho^2  \Biggr]
d \Omega_{D-2}^2.
\end{eqnarray}
We further transform the Lorentzian metric into a Euclidean metric by
a Wick rotation
\begin{equation}
\theta = i \theta_E.
\end{equation}
With the periodicity imposed
\begin{equation}
\theta_E = \frac{(D-3)(1-u)}{2 r_+} \tau_E = 2 \pi,
\end{equation}
the Euclidean metric near the event horizon  becomes topologically
$R^2 \otimes S_{D-2}$:
\begin{eqnarray}
ds^2_E &=&  \rho^2 \Biggl[1 + \frac{(D-3)}{4 r+^2}
\Bigl( (D-3) u - \frac{(D-2)(1-u)}{2} \Bigr) \rho^2 \Biggr] d \theta^2_E
\nonumber\\
&&~+ \Biggl[1 - \frac{(D-3)}{4 r+^2}
\Bigl( (D-3) u - \frac{(D-2)(1-u)}{2} \Bigr) \rho^2 \Biggr] d \rho^2
+ r_+^2 \Biggl[1 + \frac{(D-3) (1-u)}{4 r_+^2} \rho^2  \Biggr]
d \Omega_{D-2}^2.
\end{eqnarray}
This periodicity defines the Hawking temperature of the RN black hole.
\begin{equation}
\beta_H = \frac{1}{\tau_E} = \frac{4 \pi r_+}{(D-3) (1-u)}.
\label{temp}
\end{equation}
Note that the temperature can also be obtained from
\begin{equation}
\beta_H = \frac{4 \pi}{\Delta' (r_+)} .
\end{equation}
As in Sec. II, by evaluating the partition function, one finds
the Bekenstein-Hawking entropy
\begin{equation}
S_D = \frac{A_D}{4G},
\end{equation}
where
\begin{equation}
A_D = \frac{2 \pi^{\frac{D}{2}}}{\Gamma \Bigl(\frac{D}{2}\Bigr)}
r_+^{D-1}
\label{area}
\end{equation}
is the area of the event horizon.

\section{Brick Wall Model}

As an  attempt to  understand  the origin  of black  hole  entropy,
't  Hooft considered  the
statistical entropy of a scalar field outside
the event horizon of a Schwarzschild black  hole \cite{thooft}. In
order to  regulate the  ultraviolet divergence
of the Helmholtz  free energy  due to   the infinite
degeneracy of the number of states on the event horizon,
he introduced a brick wall of an
infinitesimal thickness. In this Section we apply his idea to the massive scalar field
in the $D$-dimensional RN black hole.

The massive scalar field in the RN black hole background
obeys the Klein-Gordon equation
\begin{equation}
(- g)^{-1/2} \partial_{\mu} \Bigl( g^{\mu \nu} (- g)^{1/2} \partial_{\nu}
\Phi \Bigr) - m^2 \Phi = 0.
\end{equation}
We find a stationary solution in the spherical coordinates
\begin{equation}
\Phi (x, t) = e^{- iEt} \phi (r, x_i),
\end{equation}
where $x_i, i = 2, 3,  \cdots, D$, are the coordinates  on a $(D-2)$-sphere
whose metric is related with the metric $g_S$ on the unit-sphere:
\begin{equation}
g^{i j} = \frac{1}{r^2} g^{i j}_S.
\end{equation}
Then the radial motion becomes
\begin{eqnarray}
&&\frac{1}{r^{D-2}} \frac{\partial}{\partial r} \Bigl( r^{D-2} \Delta (r)
\frac{\partial}{\partial r} \phi \Bigr) - \nonumber \\
&&~~~~~~\Bigl(\frac{l (l + D -3)}{r^2}
+ m^2 \Bigr) \phi + \frac{E^2}{\Delta (r)} \phi = 0.
\label{rad eq}
\end{eqnarray}
Eq. (\ref{rad eq}) is an analog of one-dimensional Schr\"{o}dinger equation.
We find the radial momentum $p_r = \frac{\partial {\cal S}}{\partial r}$
using the WKB wave function $\phi \propto e^{i {\cal S} (r, l, E)}$:
\begin{equation}
\hspace*{-22pt}
p_r (r, l, E) = \frac{1}{\Delta^{1/2} (r)} \Biggl[\frac{E^2}{\Delta (r)}
- \frac{l (l+D -3)}{r^2} - m^2 \Biggr]^{1/2}.
\end{equation}

The key idea of the 't Hooft brick wall is to introduce the ultraviolet and
infrared cut-offs to regulate the infinite quantities just according
to the regularization method of quantum field theory.
The scalar field is confined to a spherical shell of the inner radius $r_+ + h$ and
the outer radius $L$ and required to satisfy the Dirichlet boundary condition
\begin{equation}
\Phi ( r = r_+ + h) = 0 = \Phi(r = L).
\end{equation}
The brick wall thickness $h$ cuts off the ultraviolet divergence
and the outer radius $L$
cuts off the infrared divergence.
We are interested in the physics inside the spherical
shell. The number of radial modes is given by
\begin{equation}
n (l, E) = \frac{1}{\pi} \int_{r_+ + h}^{L} dr p_r (r, l, E),
\end{equation}
and, by summing over the angular momentum  states with
the correct degeneracy of
the angular momentum states taken into account,
we obtain the total number of states
for a given energy $E$

\begin{equation}
g (E) = \frac{1}{\pi} \int_{r_+ +h}^{L} \frac{dr}{\Delta (r)} \int dl
(2l+ D -3) \frac{(l+ D-4)!}{(D-3)! (l!)}
\Biggl[E^2 - \Bigl(m^2 + \frac{l (l+D-3)}{r^2} \Bigr) \Biggr]^{1/2}.
\end{equation}
The most divergent terms as $h$ tends to zero are found \cite{kim1}
\begin{equation}
g_{2n}^{\rm m. div} (E) = \frac{B(n-1, \frac{3}{2})}{\pi (n-1)(2n-3) (2n-3)!}
r_+^{2n-1} \Biggl( \frac{(1 - u)^2 m^2 + 2 uE^2}{(1-u)^2  m^2}\Biggr)^{n - 1/2}
\frac{E^{2n-1}}{\epsilon^{n-1}},
\end{equation}
for an even-dimensional case of $D = 2n$, and
\begin{equation}
g_{2n + 1}^{\rm m. div} (E) = \frac{1}{2^{2n-1} (n-\frac{1}{2})(2n-2)
(n-1)! n!}
r_+^{2n} \Biggl( \frac{(1 - u)^2 m^2 + 2 uE^2}{(1-u)^2  m^2}\Biggr)^{n}
\frac{E^{2n-1}}{\epsilon^{n - 1/2}},
\end{equation}
for an odd-dimensional case of $D = 2n+1$,
where $\epsilon = (D -3) \frac{h}{r_+}$ and
$u = \Bigl(\frac{r_-}{r_+} \Bigr)^{D-3}$.
The Helmholtz free energy is given by
\begin{equation}
F = - \frac{1}{\pi} \int_{0}^{\infty} dE \frac{g(E)}{e^{\beta E} -1}.
\label{helm}
\end{equation}
By making use of the integral
\begin{equation}
\int_{0}^{\infty} dx \frac{x^{n-1}}{e^{x} -1} = \Gamma (n) \zeta (n),
\end{equation}
where $\zeta$ is the Rieman zeta function,  in the large mass limit
we find the free energy  in $2n$ dimensions
\begin{equation}
 F_{2n}^{\rm m. div} =
 - \frac{B(n-1, \frac{3}{2})}{\pi^2 (n-1)(2n-3) (2n-3)!}
\frac{\Gamma (2n) \zeta(2n)}{\beta^{2n}} \frac{r_+^{2n-1}}{\epsilon^{n-1}},
\end{equation}
and in $2n+1$ dimensions
\begin{equation}
F_{2n+1}^{\rm m.div} =
- \frac{1}{\pi 2^{2n-1} (n-\frac{1}{2})(2n-2)
(n-1)! n!} \frac{\Gamma (2n) \zeta (2n)}{\beta^{2n}}
\frac{r_+^{2n}}{\epsilon^{n - 1/2}}.
\end{equation}
Finally we obtain the on-shell entropy of the scalar field
\begin{equation}
S_{2n}^{\rm m. div} =
 - \frac{2 (2n)(2n-1) \zeta(2n) B(n-1, \frac{3}{2})}{\pi^2 (2n-3)}
\Bigl(\frac{(2n-3)(1-u)}{4 \pi}\Bigr)^{2n-1} \frac{1}{\epsilon^{n-1}},
\end{equation}
in $2n$ dimensions, and
\begin{equation}
S_{2n+1}^{\rm m. div} =
- \frac{(2n)! \zeta(2n) }{\pi 2^{2n-1} (n-\frac{1}{2})(2n-2)
(n-1)! n!} \Bigl( \frac{(2n-2)(1-u)}{4 \pi} \Bigr)^{2n-1}
\frac{r_+}{\epsilon^{n - 1/2}},
\end{equation}
in $2n+1$ dimensions, respectively. By suitably choosing the brick
wall thickness $\epsilon$, the statistical entropy of the scalar field
can give rise to the Bekenstein-Hawking entropy
\begin{equation}
S_D = \frac{A_{D}}{4}.
\end{equation}
Despite of an ad hoc prescription of the brick wall,
this procedure for obtaining the black hole entropy
may have a physical origin, since the thickness has the
geometric meaning of an invariant distance from the event horizon.

\section{Renormalized One-Loop Effective Action}

From now on we shall focus on the $2n$-dimensional RN  black hole,
since the renormalization scheme was shown to work only in four \cite{demers}
and six dimensions, but not in five dimensions \cite{kim1},
and, as will be shown in the next Section, it works for all the even-dimensional
RN black holes.
So let us consider the one-loop effective action for
$2n$-dimensional gravity coupled to a massive scalar field~\cite{birrel}
\begin{equation}
{\cal I}_{2n} = \int d^{2n} x \sqrt{-g} \Bigl[- \frac{\Lambda}{8 \pi G}
+ \frac{R}{16  \pi  G} +   \frac{\alpha_1}{4 \pi} R^2   + \frac{\alpha_2}{4  \pi} R_{\mu  \nu}
R^{\mu\nu} + \frac{\alpha_3}{4 \pi} R_{\alpha \beta \mu \nu} R^{\alpha \beta \mu \nu} \Bigr]
\end{equation}
where $\Lambda_B$ is the bare cosmological constant, $G_B$ the bare gravitational constant,
and  $\alpha_{B,i}$ are the  bare   coupling  constants.
Using  the   DeWitt-Schwinger
method we find the one-loop effective action
for the massive scalar field \cite{birrel,dewitt}
\begin{equation}
W_{2n} (m) = \frac{1}{2 \bigl(4 \pi \bigr)^{n}} \int d^{2n} x \sqrt{-g} \int_0^{\infty} d(is)
\sum_{k = 0}^{\infty} a_k (x)\bigl(is \bigr)^{k - n -1} e^{- i m^2 s},
\end{equation}
where
\begin{eqnarray}
a_0 &=& 1,
~~~~a_1 = \frac{1}{6} R, \nonumber\\
a_2 &=& \frac{1}{30}  R_{,\mu}^{~~;\mu} + \frac{1}{72}  R^2
\nonumber \\
&&~~~~+ \frac{1}{180} R_{\alpha \beta
\mu \nu} R^{\alpha \beta \mu \nu} - \frac{1}{180} R_{\mu \nu}R^{\mu \nu}.
\end{eqnarray}
In order to regulate the  divergent terms of the  effective action $W_{2n}$ we  shall use the
Pauli-Villars regularization method, in which we introduce a number of
fictitious bosonic and fermionic
regulator fields of masses $m_{B_i}$ and $m_{F_i}$, respectively. Then the total action is the
sum of those of regulator and scalar fields
\begin{equation}
W_{2n} = \frac{1}{2 \bigl(4 \pi \bigr)^{n}} \int d^{2n} x \sqrt{-g} \int_0^{\infty} d(is)
\sum_{k = 0}^{\infty} a_k (x)\bigl(is  \bigr)^{k - n -  1} \bigl(\sum_i e^{- i  m_{B_i}^2 s} -
\sum_i e^{-i m_{F_i}^2 s} \bigr).
\end{equation}
Performing  the  integral  one  finds  the  divergent  contributions  to  the  effective  action
\cite{kim1}
\begin{eqnarray}
W^{{\rm div}}_{2n} = && \frac{1}{2 \bigl(4 \pi \bigr)^{n}} \int d^{2n} x \sqrt{-g}
\sum_{k = 0}^{n} a_k (x) (-1)^{n + 1 - k} \nonumber\\
&& \times \Biggl[\frac{1}{(n-k)!} \Bigl( \sum_i m_{B_i}^{2(n-k)}
\ln\bigl( m_{B_i}^2 \bigr) - \sum_i m_{F_i}^{2(n-k)} \ln\bigl( m_{F_i}^2 \bigr) \Bigr)
\nonumber\\
&& -   \frac{1}{(n-k)!} \Bigl(I_1  + \sum_{p   = 1}^{n-k}  \frac{1}{p} \Bigr)   \Bigl( \sum_i
m_{B_i}^{2(n-k)} - \sum_i m_{F_i}^{2(n-k)} \Bigr) \nonumber\\
&& + \sum_{l = 2}^{n+1-k} \frac{(-1)^l}{(n+1-k-l)!} I_{l}
\Bigl( \sum_i m_{B_i}^{2(n+1-k-l)} - \sum_i m_{F_i}^{2(n+1-k-l)} \Bigr) \Biggr],
\end{eqnarray}
where
\begin{equation}
I_p = \int_{0}^{\infty} \frac{1}{t^p}.
\label{inf int}
\end{equation}
To remove the  infinite constants  $I_p$ given by  Eq. (\ref{inf  int}), we impose  the mass
conditions
\begin{equation}
\sum_i m_{B_i}^{2(n-k)} = \sum_i m_{F_i}^{2(n-k)}
\label{mass con}
\end{equation}
for $k = 0, 1, \cdots, n$. We are then left with the renormalized action
\begin{equation}
W^{{\rm ren}}_{2n} = \int d^{2n} x \sqrt{-g} \sum_{k = 0}^{n} a_k (x)
\frac{{\cal B}_k}{2 \bigl(4 \pi \bigr)^{n} (n-k)!},
\end{equation}
where
\begin{eqnarray}
{\cal B}_k &=& (-1)^{n + 1 - k} \Bigl(
\sum_i m_{B_i}^{2(n-k)} \ln\bigl( m_{B_i}^2 \bigr)
\nonumber \\
&&~~~~~~~~- \sum_i m_{F_i}^{2(n-k)}
\ln\bigl( m_{F_i}^2 \bigr) \Bigr)
\label{ren con1}
\end{eqnarray}
are the renormalization constants. We may now
renormalize the one-loop effective action for
gravity plus matter field by  redefining the cosmological,  gravitational, and coupling
constants
\begin{eqnarray}
&& \frac{\Lambda}{8\pi G} - \frac{{\cal B}_0}{2 (4\pi)^n n!} =
\frac{\Lambda^{\rm ren}}{8\pi G^{\rm ren}}, \nonumber\\
&& \frac{1}{16\pi G} + \frac{{\cal B}_1}{12 (4\pi)^n (n-1)!} =
\frac{1}{16\pi G^{\rm ren}}, \nonumber\\
&& \frac{\alpha_1}{4\pi} + \frac{{\cal B}_2}{144 (4\pi)^n (n-2)!} =
\frac{\alpha^{\rm ren}_1}{4\pi}, \nonumber\\
&& \frac{\alpha_2}{4\pi} - \frac{{\cal B}_2}{360 (4\pi)^n (n-2)!} =
\frac{\alpha^{\rm ren}_2}{4\pi}, \nonumber\\
&& \frac{\alpha_3}{4\pi} + \frac{{\cal B}_2}{360 (4\pi)^n (n-2)!} =
\frac{\alpha^{\rm ren}_3}{4\pi}.
\label{ren}
\end{eqnarray}

\section{Renormalization of RN Black Holes in $2n$-dimensions}

We shall now compute the statistical entropy of the massive scalar field in the $2n$-dimensional
RN black  hole background.  The statistical  entropy of  the  scalar field  in the  RN black  hole
background has already  been found  in four \cite{demers} and   six dimensions
\cite{kim1}.
By applying the  WKB idea  to the quantum  mechanical
Klein-Gordon equation  we are able  to find  the
number of states, $g(E)$, for a given energy and to
obtain the Helmholtz free energy (\ref{helm}).

By taking  the degeneracy  of angular  momentum states   for a fixed  $l$, the  number of
states of the massive scalar field is given by
\begin{equation}
g_{2n} (E,m) = \frac{1}{\pi} \int_{r_+ + h}^{L} \frac{dr}{\Delta(r)} \int dl (2l+2n-3)
\frac{(l+ 2n -4)!}{(2n-3)! l!} \sqrt{E^2 - \Bigl(m^2 + \frac{l(l+2n-3)}{r^2}
\Bigr) \Delta(r)},
\label{num st}
\end{equation}
where $L$ is an infrared  cutoff and $h$ is a  brick wall introduced to  regulate the possible
ultra-violet divergence. Following Ref. \cite{kim1} we obtain the number of states
\begin{eqnarray}
g_{2n} (E,m) =  \frac{1}{\pi (2n-3)!} \sum_{k = 0}^{n-2}  C_k^{2n}
B \Bigl(k+1, \frac{3}{2} \Bigr) \frac{r_+^{2k+3}}{2n-3}
&&   \times   \int_{\epsilon}    dx   \frac{   \Bigl[E^2   -    x(1-u+ux)m^2   \Bigr]^{k   +
\frac{3}{2}}}{(1-x)^{1+\frac{2k+3}{2n-3}} x^{k+2}(1-u+ux)^{k+2}},
\label{f ns}
\end{eqnarray}
where
\begin{eqnarray}
x &=& 1 - \Bigl(\frac{r_+}{r} \Bigr)^{2n-3}, \nonumber\\
\epsilon &=& (2n-3)\frac{h}{r_+}, \nonumber\\
u &=& \Bigl(\frac{r_-}{r_+} \Bigr)^{2n-3}.
\end{eqnarray}
As the  explicit calculation in four and six dimensions  shows,
there are two typical types of  terms: the terms  to be removed
by the same mass  conditions (\ref{mass con})
and the terms  contributing to  the free energy  and entropy,
which are later on to  be
renormalized by redefining the coupling constants (\ref{ren}).
We are, however, interested  in the contribution to  the
entropy proportional to  the area of  the event horizon  after removing all the  possible divergent
terms which depend on the infinitely large masses of regulator fields.

After some gymnastic of algebra, we get the number of states \cite{kim1}
\begin{eqnarray}
g_{2n} (E,m) = && \frac{1}{\pi (2n-3)!} \sum_{k = 0}^{n-2}  C_k^{2n}
B \Bigl(k+1, \frac{3}{2} \Bigr) \frac{r_+^{2k+3}}{2n-3} \sum_{q = 0}^{\infty} H_q^{2n,k}
\sum_{p = 0}^{\infty} (-1)^p {{k+ \frac{3}{2}} \choose {p}} \nonumber\\
&& \times \Bigl(\frac{(1-u)^2 m^2 + 2uE^2}{ (1-u)^2 m^2} \Bigr)^{k+ \frac{3}{2}}
\Bigl( \frac{u}{(1-u)^2 m^2 + 2uE^2}\Bigr)^{p}
\sum_{l = 0}^{p} {p \choose l} E^{4l} \int_{\epsilon} dx
x^{-k-2 + q} Z^{k +p -2l + \frac{3}{2}},
\end{eqnarray}
where $H_q^{2n,k}$ are the coefficients of Taylor expansion
\begin{equation}
\frac{1}{(1 -x)^{1 + \frac{2k + 3}{2n-3}} (1 - u + ux)^{k+ 2}} = \sum_{q=0}^{\infty}
H_q^{2n,k} x^q,
\end{equation}
and
\begin{equation}
Z = E^2 - (1 - u) m^2 x.
\end{equation}
A careful scrutiny shows that the term with $k = n-2, q = 0, p = 0, l = 0$
\begin{equation}
g_{2n}^{\rm area} (E,m) =  \frac{C_{n-2}^{2n}}{\pi \Gamma(2n-2)} B  \Bigl(n - 1, \frac{3}{2}
\Bigr) \frac{r_+^{2n-1}}{2n-3} H_0^{2n,n-2} \int_{\epsilon} dx \frac{Z^{n - \frac{1}{2}}}{x^n}
\end{equation}
contributes to the entropy proportional to the area of event horizon.
It is found that
\begin{equation}
H_{0}^{2n, n-2} = \frac{1}{(1-u)^n},~ C^{2n}_{n-2} = 1.
\end{equation}
We now compute the integral
\begin{eqnarray}
A^{n}_{n - \frac{1}{2}} &=& \int_{\epsilon} dx \frac{ Z^{n - \frac{1}{2}}}{x^n}
\nonumber\\
&=& Z^{n + \frac{1}{2}} (\epsilon) \Biggl[ \frac{1}{(n -1) E^2}\frac{1}{\epsilon^{n -1}}
 + \sum_{l = 1}^{n -2}  (-1)^l \frac{1 \cdot 3  \cdot 5 \cdots (2l+1)}{  2^l (n-1) (n-2) \cdots
(n-l-1)}
\frac{1}{\epsilon^{n-l-1}} \frac{1}{E^2} \Bigl( \frac{(1-u)m^2}{E^2} \Bigr)^{l}
\Biggr] \nonumber\\
&& + ~ (-1)^{n} \frac{1 \cdot 3 \cdot 5  \cdots (2n-1)}{ 2^{n-1} (n-1)!}
\Bigl(\frac{(1-u)m^2}{E^2} \Bigr)^{n-1}
\Biggl[ - 2 \sum_{l =1}^{n} \frac{E^{2n - 2l}}{2l-1} Z^{l- \frac{1}{2}} (\epsilon)
- E^{2n-1} \ln \Bigl(\frac{E - Z^{\frac{1}{2}} (\epsilon)}{ E + Z^{\frac{1}{2}} (\epsilon)} \Bigr)
\Biggr].
\end{eqnarray}
Note that the term proportional to $m^{2n-2} E$ is
\begin{equation}
A_{n - \frac{1}{2}}^{n} = (1)^{n+1} \frac{(2n-1)!!}{2^{n-1} (n-1)!}
\Bigl[(1-u)m^2 \Bigr]^{n-1} E \ln \Bigl(\frac{(1-u)m^2}{4 E^2} \epsilon \Bigr)
\end{equation}
and that
\begin{equation}
B(n-1,  \frac{3}{2})  =  \frac{\pi^{1/2}  \Gamma  (n-1)}{  2  (n-  \frac{1}{2})  \Gamma(n  -
\frac{1}{2})}.
\end{equation}
So the number of states  becomes
\begin{equation}
g_{2n}^{\rm area} (E, m) = \frac{(-1)^{n+1}}{2^{2n-1} \pi^{1/2} (n-1)! \Gamma (n - \frac{1}{2})}
\frac{r_{+}^{2n-1}}{(2n-3) (1-u)} m^{2n-2} E \ln \Bigl(\frac{(1-u)m^2}{4 E^2} \epsilon \Bigr).
\end{equation}
Finally we obtain the off-shell free energy
\begin{equation}
F_{2n}  (m)  =   \frac{(-1)^{n+1} \pi^{3/2}}{12   2^{2n-2}  \pi^{1/2}  (n-1)!  \Gamma   (n -
\frac{1}{2})} \frac{r_{+}^{2n-1}}{(2n-3) (1-u) \beta^2} m^{2n-2} \ln (m^2 \epsilon),
\end{equation}
and entropy
\begin{equation}
S_{2n} (m) = \frac{(-1)^{n} \pi^{3/2}}{12 2^{2n-3} \pi^{1/2} (n-1)! \Gamma (n - \frac{1}{2})}
\frac{r_{+}^{2n-1}}{(2n-3) (1-u) \beta} m^{2n-2} \ln (m^2 \epsilon).
\end{equation}
Substituting the Hawking temperature (\ref{temp}) and
the area of event horizon  (\ref{area}),
we find the on-shell entropy of the scalar field
\begin{eqnarray}
\hspace*{-15pt}
S_{2n} (m) &=& \frac{A_{2n} }{4} \frac{(-1)^{n}}
{12 (4 \pi)^{n-1} (n-1)!} m^{2n-2} \ln (m^2 \epsilon).
\end{eqnarray}

We introduce the same number of fictitious bosonic  or fermionic regulator  fields which obey  the same Bose-Einstein
distribution but   with opposite  signs~\cite{demers}. To   remove ultraviolet  divergence  of
entropy, we  use the  Pauli-Villars regularization   method. Let the  masses of  bosonic  and
fermionic regulator fields  be $m_{B_i}$ and $m_{F_i}$,  respectively. Thus the total entropy
is the sum of each field
\begin{equation}
S_{2n} = \sum_{i} S_{2n} (m_{B_i}) - \sum_{i} S_{2n} (m_{F_i}).
\end{equation}
The divergence coming from the brick wall thickness is removed from  the mass condition in
Sec. V. In the end we obtain the renormalized entropy of the scalar field
\begin{equation}
S_{2n} = \frac{A_{2n} }{4} \frac{{\cal B}_{1}}{12 (4 \pi)^{n-1}}.
\end{equation}
Thus we are able to show that the total entropy of the Bekenstein-Hawking entropy and
the statistical entropy of the scalar
field can be renormalized and lead to the correct renormalized
black hole area-law
\begin{equation}
S_{2n}^{\rm ren} = \frac{A_{2n}}{4G}   +   \frac{A_{2n}    }{4}
\frac{{\cal   B}_{1}}{12   (4    \pi)^{n-1}}   =
\frac{A_{2n}}{4G^{\rm ren}}.
\label{ren ent}
\end{equation}
Equation (\ref{ren ent}) is the main result of this talk.

\section{Discussion and Conclusion}

In this talk we revisited the renormalization of black hole entropy.
It was found that only for the even dimensional Reissner-N\"{o}rdstrom black holes
the total entropy of the black hole's geometric entropy and the statistical entropy
of a massive scalar field
gives the correct Bekenstein-Hawking black hole entropy
in terms of the renormalized gravitational constant, which is the same
as that appeared in the one-loop effective action for gravity minimally
coupled to the massive scalar field.
The covariant Pauli-Villars regularization method used to regularize both the
one-loop effective action and the statistical entropy of the scalar field
makes use of the fictitious  regulator fields that obey the same Bose-Einstein
distribution but contribute to the Helmholtz free energy with opposite signs.
Even though not shown in this talk, this suggests that a black hole coupled to
both bosonic scalar fields and
fermionic fields may have the regularized black hole entropy. This implies
that supersymmetry may play a role in understanding the microscopic origin of
black hole entropy.

\section*{acknowledgments}
This work was supported in parts by the Non-Directed Research Fund,
Korea Research Foundation, 1996, and by the Basic Science Research
Institute Program, Korea Ministry of Education under Projects Nos.
BSRI-97-2418, BSRI-97-2425, and BSRI-97-2427.

\end{document}